\documentclass[trackchanges]{aastex701}
\usepackage{amsmath}
\usepackage{graphicx}

\begin{document}

\title{A Tale of Two Jets: Double Relativistic Outflows from Close Binary GRB Progenitors}

\author[orcid=0000-0003-2516-6288,sname='Gao']{He Gao}
\affiliation{
Institute for Frontier in Astronomy and Astrophysics, Beijing Normal University, Beijing 102206, China}
\affiliation{School of Physics and Astronomy, Beijing Normal University, Beijing 100875, China}
\email[show]{gaohe@bnu.edu.cn} 

\correspondingauthor{He Gao}

\begin{abstract}
Gravitational wave astronomy has revealed that close binaries with compact companions are widespread. Long GRBs (LGRBs) from massive star collapse face persistent challenges in achieving the rapid core rotation required by the collapsar model. Binary interaction via tidal spin-up offers a natural solution; recent population synthesis studies suggest a substantial fraction of LGRBs may originate from close binaries with a compact companion. In this scenario, supernova ejecta from the primary can be accreted by the companion, potentially launching a second relativistic jet after a delay set by the binary separation. We develop a comprehensive model for these double-jet systems, analyzing the dynamics of the second jet and its interaction with the first. The resulting observational signatures depend critically on the Lorentz factor ratio, the alignment angle, and the time delay. For aligned jets, two regimes arise: a fast second jet producing multiple gamma-ray triggers with distinct spectral/polarization evolution, and a slow second jet where its emission appears as an X-ray flare followed by an afterglow plateau from energy injection. For misaligned jets, the observed signal ranges from normal GRBs with late-time radio structures to fast X-ray transients followed by off-axis rebrightening. These features have observational parallels in existing GRB data. High-resolution radio interferometry with SKA, time-resolved polarimetry with eXTP, and multi-wavelength surveys with Einstein Probe and SVOM will test these predictions, providing constraints on the evolution of close massive binaries as progenitors of GRBs and gravitational wave sources.
\end{abstract}


\section{Introduction}

The dawn of gravitational wave astronomy, marked by the first direct detection by Advanced LIGO in 2015 \citep{abbott2016observation}, has opened a new window for understanding compact binary systems. To date, the LIGO-Virgo-KAGRA collaboration has confidently detected over 200 mergers of compact object binaries \citep{abbott2021gwtc2, 2023PhRvX..13d1039A}, including binary black holes, binary neutron stars, and neutron star-black hole systems. These discoveries have revolutionized our understanding of the formation channels and evolutionary pathways of close binaries \citep{mandel2022mergers}, revealing that such systems are widespread in the universe and can involve a diverse range of compact object masses and spins \citep{2023PhRvX..13a1048A}.

Coincidentally, gamma-ray bursts (GRBs) have long been theorized to originate from similar compact object systems. Short GRBs are firmly associated with neutron star mergers \citep[e.g.,][]{eichler1989, narayan1992}, a connection spectacularly confirmed by the multi-messenger event GW170817 \citep{abbott2017grb}.The core collapse of massive stars is the traditional progenitor model for long GRBs \citep{woosley1993, macfadyen1999}. However, this model faces a persistent challenge: efficient angular momentum transport in stellar interiors tends to spin down the cores of isolated massive stars, making it difficult to achieve the rapid rotation required for jet launching \citep{spruit2002, heger2005, fuller19}. Binary interaction offers a natural solution, as tidal forces in close binaries can spin up the primary star's core to the degree necessary to launch a relativistic jet within the collapsar framework \citep{cantiello2007, vandenheuvel07, detmers2008, qin18}. Recent binary population synthesis studies calibrated with both gravitational wave and GRB observations suggest that a substantial fraction of long GRBs may indeed originate from close binary systems, with estimates ranging from approximately 20\% to 85\% \citep{chrimes20,bavera22}.

In such binaries, the companion is often a compact object left behind by an earlier supernova \citep[e.g.,][]{bethe1998, podsiadlowski2003}. When the primary finally explodes, its ejecta can be gravitationally captured by the compact companion, forming an accretion disk that may launch a second relativistic jet after a time delay determined by the binary separation \citep{gao20,soker20}. \citet{gao20} explored the supernova signatures arising from such systems, showing that accretion feedback from the companion can produce distinctive features in the supernova light curve. 

In this paper, we develop a comprehensive model for double-jet systems arising from close binaries with compact companions, with the goal of predicting the observable signatures of such interactions and identifying key features that can be tested with current and upcoming multi-wavelength observations.

\section{Model Description}
\label{Sec:Model}

\subsection{Binary System and Ejecta Profile}

We consider a close binary system consisting of a massive star (the primary) and a compact companion. We first consider the case where the compact companion is a black hole (BH) of mass $M_{\mathrm{BH}}$. The scenario with a neutron star (NS) companion will be discussed later. 

In such a close binary, tidal interactions with the compact companion can spin up the core of the primary star. Recent binary population synthesis studies \citep{bavera22} have demonstrated that this tidal spin-up is efficient enough to produce rapidly rotating cores at collapse, a key requirement for launching a relativistic jet under the collapsar model \citep{woosley1993}. Consequently, the primary core collapse gives rise to a highly relativistic jet, hereafter referred to as the first jet (J1). This jet is responsible for producing the observed long gamma-ray burst (LGRB). Following the standard picture \citep[e.g.,][]{zhang18book}, J1 is characterized by an isotropic-equivalent luminosity $L_{\mathrm{iso,1}} \sim 10^{50} - 10^{52}~\mathrm{erg~s^{-1}}$, a Lorentz factor $\Gamma_1 \sim 100 - 300$, and a half-opening angle $\theta_j \sim 0.05 - 0.1$ rad. This jet successfully breaks out of the stellar envelope, producing the observed prompt gamma-ray emission and subsequently driving the multi-wavelength afterglow through its interaction with the circumstellar medium.

When the primary explodes as a supernova (SN), a total mass $M_{\mathrm{ej}}$ is ejected with an explosion energy $E_{\mathrm{sn}}$. Based on numerical simulations of SN explosions, the density profile of the SN ejecta can be described by a broken power law \citep{matzner99}:
\begin{equation}
  \rho_{\mathrm{ej}} (v, t) = \left\{ \begin{array}{ll}
    \zeta_{\rho} \dfrac{M_{\mathrm{ej}}}{v_{\mathrm{tr}}^3 t^3} \left(
    \dfrac{r}{v_{\mathrm{tr}} t} \right)^{- \delta}, & v_{\mathrm{ej}, \min}
    \leqslant v < v_{\mathrm{tr}}\\[10pt]
    \zeta_{\rho} \dfrac{M_{\mathrm{ej}}}{v_{\mathrm{tr}}^3 t^3} \left(
    \dfrac{r}{v_{\mathrm{tr}} t} \right)^{- n}, & v_{\mathrm{tr}} \leqslant v
    \leqslant v_{\mathrm{ej}, \max}
  \end{array} \right.
\end{equation}
where the transition velocity $v_{\mathrm{tr}}$ is obtained from the density continuity condition:
\begin{equation}
\begin{aligned}
  v_{\mathrm{tr}} = \zeta_v \left( \frac{E_{\mathrm{sn}}}{M_{\mathrm{ej}}}
  \right)^{1 / 2} 
  \simeq 1.2 \times 10^4 \ \mathrm{km \ s^{-1}} \left(
  \frac{E_{\mathrm{sn}}}{10^{51} \ \mathrm{erg}} \right)^{1 / 2} \left(
  \frac{M_{\mathrm{ej}}}{M_{\odot}} \right)^{- 1 / 2}.
\end{aligned}
\end{equation}
The numerical coefficients depend on the density power indices \citep{kasen16}:
\begin{equation}
  \zeta_{\rho} = \frac{(n - 3) (3 - \delta)}{4 \pi (n - \delta)}, \qquad
  \zeta_v = \left[ \frac{2 (5 - \delta) (n - 5)}{(n - 3) (3 - \delta)}
  \right]^{1 / 2}.
\end{equation}
For core-collapse SNe, typical values are $\delta = 1$, $n = 10$ \citep{chevalier89}. The ejecta undergo homologous expansion, i.e., $r = v t$, with minimum and maximum velocities $v_{\mathrm{ej}, \min}$ and $v_{\mathrm{ej}, \max}$, and corresponding initial radii $R_{\min,0}$, $R_{\max,0}$ at the onset of the homologous phase.

\begin{figure}
\centering
\label{fig}
   \includegraphics[width=7in]{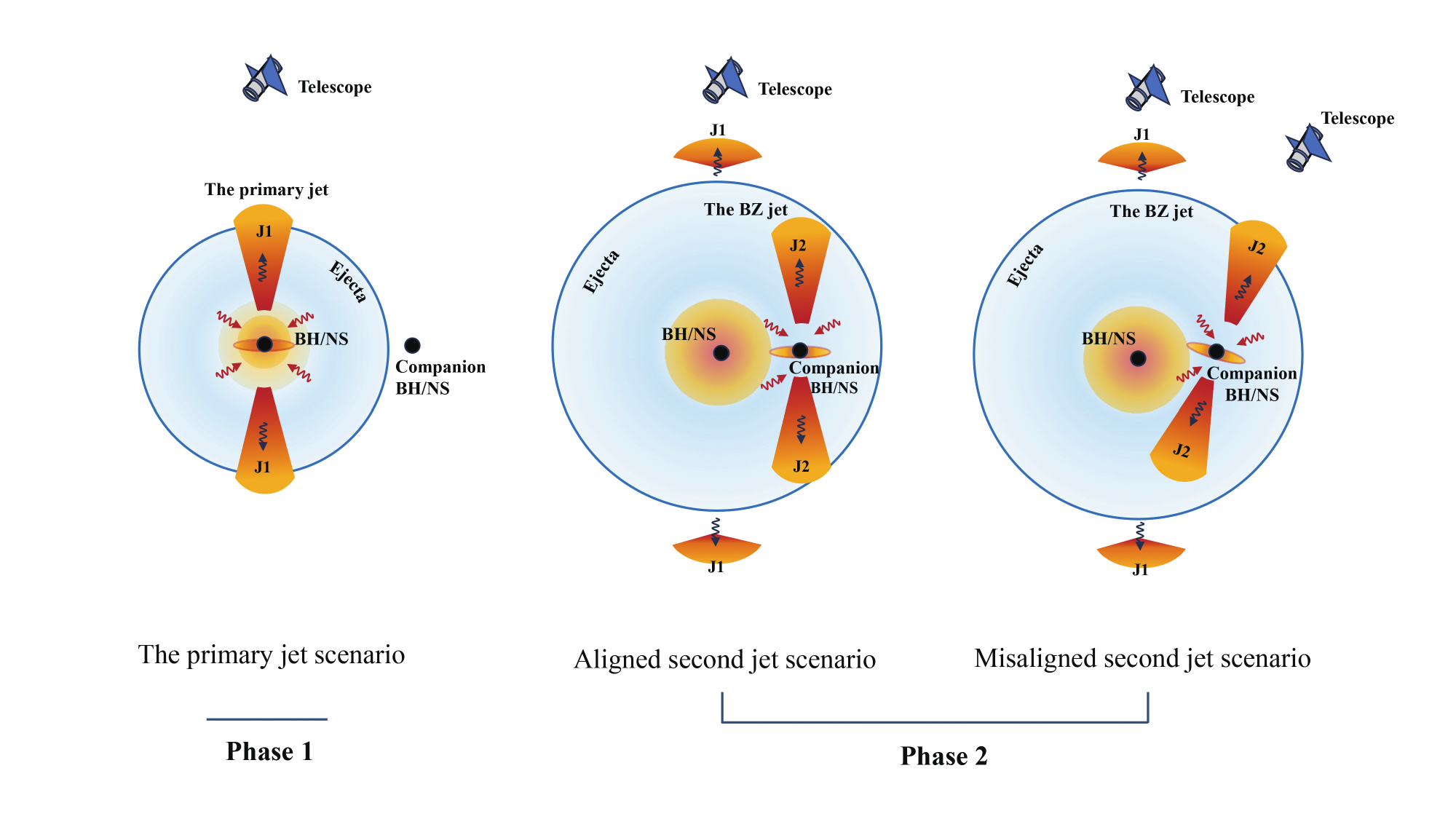}
      \caption{Schematic illustration of the two-jet scenario. Phase I: the primary jet J1 produces prompt gamma-ray emission. Phase II: the second jet J2 is launched after a time delay $(\Delta t)$. For aligned jets (left), the relative speed determines whether J2 collides violently ($\Gamma_2>\Gamma_1$) or merges gradually ($\Gamma_2<\Gamma_1$). For misaligned jets (right), the line of sight determines the observed signal: a normal GRB with late-time radio lobes if aligned with J1, or a fast X-ray transient followed by off-axis rebrightening if aligned with J2.}
            \end{figure}

\subsection{Accretion by the Companion Black Hole}

As the SN ejecta expand, a portion of the material is gravitationally captured by the companion BH. The infall begins when the outermost ejecta layer reaches $d - R_{\mathrm{acc}}$, at a time
\begin{equation}
t_{\mathrm{delay}} = \frac{d - R_{\mathrm{acc}} - R_{\max,0}}{v_{\mathrm{ej}, \max}}
\approx 10^{3} \ \mathrm{s} \left(\frac{d}{10^{12} \ \mathrm{cm}}\right)
\left(\frac{v_{\mathrm{ej}, \max}}{10^4 \ \mathrm{km \ s^{-1}}}\right)^{-1}.
\end{equation}
where $d$ is the orbital separation, $R_{\mathrm{acc}} = 2GM_{\mathrm{BH}}/v^2$ defines the boundary within which an ejecta element is gravitationally bound. 

During this phase, the mass capture rate is
\begin{equation}
  \dot{M} \simeq \pi R_{\mathrm{acc}}^2 v \rho_{\mathrm{ej}}
  = \frac{4 \pi G^2 M_{\mathrm{BH}}^2}{d^3} \zeta_{\rho}
  \frac{M_{\mathrm{ej}}}{v_{\mathrm{tr}}^3} \left( \frac{d}{v_{\mathrm{tr}} t}
  \right)^{- n}, \quad t_{\mathrm{delay}} \leqslant t < t_{\mathrm{tr}},
\end{equation}
where $t_{\mathrm{tr}} \sim d / v_{\mathrm{tr}}$ is the time when the falling region reaches the inner ejecta (i.e., when the velocity of the infalling material becomes $v_{\mathrm{tr}}$). At this characteristic time,
\begin{equation}
  \dot{M}_{\mathrm{tr}} \simeq 4.1 \times 10^{-6} M_{\odot} \ \mathrm{s^{-1}}
  \left( \frac{M_{\mathrm{ej}}}{10M_{\odot}} \right)^{5/2} \left(
  \frac{M_{\mathrm{BH}}}{20 M_{\odot}} \right)^2
  \left( \frac{d}{10^{12} \ \mathrm{cm}} \right)^{-3}
  \left( \frac{E_{\mathrm{sn}}}{10^{51} \ \mathrm{erg}} \right)^{-3/2}.
  \label{mff}
\end{equation}
For $t > t_{\mathrm{tr}}$, the density structure follows $\rho \propto r^{-\delta}$, so the fallback rate becomes
\begin{equation}
  \dot{M} = \dot{M}_{\mathrm{tr}} \left( \frac{t}{t_{\mathrm{tr}}} \right)^{\delta}, \quad t_{\mathrm{tr}} \leqslant t \leqslant t_{\mathrm{end}},
\end{equation}
where $t_{\mathrm{end}} = \frac{d - R_{\mathrm{acc}} - R_{\min,0}}{v_{\mathrm{ej}, \min}}
\sim 10^{4} \ \mathrm{s} \left(\frac{d}{10^{12} \ \mathrm{cm}}\right)
\left(\frac{v_{\mathrm{ej}, \max}}{10^3 \ \mathrm{km \ s^{-1}}}\right)^{-1}$ marking the end of the steady mass capture phase. After $t_{\mathrm{end}}$, the tail of the mass capture rate could be modeled as an exponential cutoff $\dot{M} = \dot{M}_{\mathrm{tr}} (t_{\mathrm{end}}/t_{\mathrm{tr}})^{\delta} e^{-(t-t_{\mathrm{end}})/t_{\mathrm{end}}}$.

{Once captured, the material first falls freely from the accretion radius $R_{\mathrm{acc}}$ to the circularization radius $R_{\mathrm{circ}}$ (about 10–100 gravitational radii of the BH for typical parameters), where an accretion disk forms. The free-fall time from $R_{\mathrm{acc}}$ to $R_{\mathrm{circ}}$ is $t_{\mathrm{ff,1}} \approx 0.68 \mathrm{s}(R_{\mathrm{acc}}/10^{9} \mathrm{cm})^{3/2} (M_{\mathrm{BH}}/20 M_{\odot})^{-1/2}$, which sets the mass supply timescale. Within the disk, the viscous accretion timescale is $t_{\mathrm{acc}} \sim t_{\mathrm{ff,2}}/\alpha$, where $t_{\mathrm{ff,2}} \sim (R_{\mathrm{circ}}^3/GM_{\mathrm{BH}})^{1/2} \sim 10^{-3}\,\mathrm{s}$ is the free-fall time from $R_{\mathrm{circ}}$ and $\alpha \sim 0.1$ is the viscosity parameter \citep{shakura73}. Hence the disk can quickly transport material to the BH compared to the rate at which new material is supplied. We therefore adopt the fast accretion approximation $\dot{M}_{\mathrm{acc}} \approx \dot{M}$.}

\subsection{Second Jet Launch and Power}

For a BH of mass $M_{\mathrm{BH}}$, the Eddington accretion rate is defined as $\dot{M}_{\mathrm{Edd}} = L_{\mathrm{Edd}} / (\eta c^2)$, where $L_{\mathrm{Edd}} \approx 1.3 \times 10^{38} (M_{\mathrm{BH}}/M_{\odot})\,\mathrm{erg\,s^{-1}}$ is the Eddington luminosity and $\eta \sim 0.1$ is the typical radiative efficiency. For $M_{\mathrm{BH}} = 20 M_{\odot}$, this yields $\dot{M}_{\mathrm{Edd}} \sim 2.5 \times 10^{-7} M_{\odot}\,\mathrm{yr^{-1}} \sim 8 \times 10^{-15} M_{\odot}\,\mathrm{s^{-1}}$. As shown in Eq.~(\ref{mff}), the typical fallback rate onto the companion can reach $\dot{M}_{\mathrm{tr}} \sim 4 \times 10^{-6} M_{\odot}\,\mathrm{s^{-1}}$ for typical parameters, corresponding to an accretion rate $\sim 5 \times 10^{8}$ times the Eddington rate. Such extreme super-Eddington accretion inevitably leads to powerful outflows and jets. Two main jet production mechanisms are possible: the Blandford–Znajek (BZ) mechanism \citep{BZ} extracting BH spin energy, and the Blandford–Payne (BP) mechanism \citep{BP} extracting disk rotational energy. For the BZ mechanism to operate efficiently, the BH must possess significant spin. In the context of a binary system, the companion BH is not formed in isolation; prior to the primary's explosion, it can acquire significant angular momentum through tidal interactions with the primary and accretion from the primary's stellar wind. This pre-existing spin, combined with the subsequent hyperaccretion of supernova ejecta, ensures that the companion BH possesses sufficient angular momentum to launch a powerful BZ jet. 

The BZ jet power can be estimated as \citep{lee00, li00, wang02, mckinney05, lei11, lei13, liu17}
\begin{equation}
L_{\mathrm{BZ}} = 1.7 \times 10^{50} a^2 \left( \frac{M_{\mathrm{BH}}}{M_{\odot}} \right)^2
B_{\mathrm{H},15}^2 F(a) \ \mathrm{erg \ s^{-1}},
\label{eq_Lmag}
\end{equation}
where $a$ is the BH spin parameter, $F(a)=[(1+q^2)/q^2][(q+1/q) \arctan q-1]$ with $q = a/(1+\sqrt{1-a^2})$, and $B_{\mathrm{H}}$ is the magnetic field on the horizon. Equating magnetic pressure to the ram pressure at the inner edge gives
\begin{equation}
\frac{B_{\mathrm{H}}^2}{8\pi} = P_{\mathrm{ram}} \sim \rho c^2 \sim \frac{\dot{M}_{\mathrm{acc}} c}{4\pi r_{\mathrm{H}}^2},
\label{Bmdot}
\end{equation}
with $r_{\mathrm{H}} = (1+\sqrt{1-a^2}) r_{\mathrm{g}}$, $r_{\mathrm{g}} = G M_{\mathrm{BH}} / c^2$. Combining, we obtain a time-dependent BZ luminosity:
\begin{equation}
  L_{\mathrm{BZ}}(t) = \eta_{\mathrm{BZ}} \dot{M}_{\mathrm{tr}} c^2 \left\{
  \begin{array}{ll}
    (t/t_{\mathrm{tr}})^{n}, & t_{\mathrm{delay}} \leqslant t < t_{\mathrm{tr}}\\[4pt]
    (t/t_{\mathrm{tr}})^{\delta}, & t_{\mathrm{tr}} \leqslant t \leqslant t_{\mathrm{end}}\\[4pt]
    (t_{\mathrm{end}}/t_{\mathrm{tr}})^{\delta} e^{-(t-t_{\mathrm{end}})/t_{\mathrm{end}}}, &  t > t_{\mathrm{end}}
  \end{array} \right.
\end{equation}
where $\eta_{\mathrm{BZ}} = 0.52 a^2 F(a)/(1+\sqrt{1-a^2})^2$ (e.g., $\eta_{\mathrm{BZ}} \approx 0.0008$ for $a=0.1$, and $0.17$ for $a=0.9$).

The BP outflow luminosity would have the same temporal shape as $L_{\mathrm{BZ}}$ but with efficiency \citep{armitage99}
\begin{equation}
\eta_{\mathrm{BP}} = \frac{1}{16} \left( 1 + \sqrt{1 - a^2} \right)^{1/2}
\frac{R_{\mathrm{ms}}^{3/2}}{(R_{\mathrm{ms}}^{3/2} + a)^2},
\end{equation}
which is $\sim 0.006$ for $a=0.1$ and $\sim 0.013$ for $a=0.9$. Here $R_{\mathrm{ms}}$ marks the marginally stable orbit radius.


The key difference between these two mechanisms lies in their collimation and ability to penetrate the overlying SN ejecta. The BZ mechanism produces a highly relativistic, well-collimated jet. When the SN envelope expands to $R_{\rm SN}$, the time required for this jet to break out can be roughly estimated as \citep{bromberg11}:
\begin{equation}
t_{B}\approx 3000s\times L_{\rm BZ,45}^{-1/3}\theta_{10^{\rm o}}^{4/3}R_{\rm SN,13}^{2/3}M_{\rm ej,10\odot}^{1/3}.
\end{equation}
For typical parameters characterizing the companion BH's accretion system, this breakout timescale is shorter than the duration of the accretion process, allowing the BZ jet to successfully emerge from the SN envelope. 

In contrast, the BP outflow is much less collimated and has a lower characteristic velocity. Such wide-angle outflows are generally unable to break through the dense SN ejecta; instead, they are expected to be shock-heated and deposit most of their energy into the expanding envelope \footnote{Such an interaction was simulated for an NS companion by \cite{2020ApJ...901...53A}}. In this scenario, the SN bolometric luminosity can be expressed by \citep{arnett82}:
\begin{equation}
  L_{\mathrm{SN}} (t) = e^{- \left( \frac{t^2}{\tau_m^2} \right)} \int_0^t 2
  \frac{t}{\tau_m^2} L_{\mathrm{heat}} (t') e^{\left( \frac{t^{\prime
  2}}{\tau_m^2} \right)} d t',
\end{equation}
where 
\begin{equation}
  \tau_m = \left( \frac{2 \kappa M_{\mathrm{ej}}}{\beta v c} \right)^{1 / 2}
\end{equation}
is the effective diffusion timescale, where $\kappa$ is the opacity of the SN ejecta, and $\beta = 13.8$ is a constant for the density distribution of the ejecta. 
\begin{equation}
  L_{\mathrm{heat}} (t) \approx L_{\mathrm{BP}} (t) + L_{\mathrm{Ni}} (t),
\end{equation}
where $L_{\mathrm{Ni}}$ is the heating power from the radioactive decay of $^{56}$Ni. When $L_{\mathrm{BP}}$ dominates over $L_{\mathrm{Ni}}$, the resulting light curve can exhibit either a sharp early peak with luminosity reaching $L_{\mathrm{peak}} \sim 10^{44}~\mathrm{erg~s^{-1}}$ (comparable to superluminous supernovae) or an extended plateau phase. The specific morphology depends on the relative strength of $L_{\mathrm{BP}}$ compared to the nickel heating, as well as the temporal evolution of the BP outflow (see details in \cite{gao20}). 

The emergence of the BZ jet from the companion system marks the launch of a second relativistic outflow, which we denote as J2. The presence of this second jet fundamentally distinguishes the GRB signal from that of a standard single-jet event. In close binary systems, the observed gamma-ray emission consists of two components: the prompt radiation from J1 (the primary jet) and a delayed contribution from J2, superimposed after a time interval $\Delta t_{\mathrm{obs}}$. 
This superposition can produce complex light curve morphologies, including double-peaked structures or extended emission. Moreover, physical interactions between the two jets, such as collisions (if $\Gamma_2 > \Gamma_1$) or tail impacts (if $\Gamma_2 < \Gamma_1$), can generate additional distinctive signals, including delayed X-ray flares, peculiar spectral evolution, and possibly unique radio morphologies. These observational consequences will be explored in detail in the following sections.

We now turn to the case where the compact companion is a neutron star (NS). In this case, the NS accretes material from the primary ejecta, forming a hyperaccretion disk around it. For typical parameters, the accretion rate onto the companion can reach $\dot{M}_{\mathrm{tr}} \sim 10^{-6} M_{\odot}\,\mathrm{s^{-1}}$. Even at this extreme rate, the total accreted mass over the characteristic accretion timescale $t_{\mathrm{acc}} \sim 10^3$-$10^4$ s is only $\Delta M \sim \dot{M}_{\mathrm{tr}} t_{\mathrm{acc}} \sim  10^{-3}$-$10^{-2} M_{\odot}$. This is far below the $\sim 0.5$-$1.5 M_{\odot}$ needed to push a typical neutron star ($M_{\mathrm{NS}} \sim 1.4 M_{\odot}$) over the maximum stable mass ($M_{\mathrm{max}} \sim 2$-$3 M_{\odot}$, depending on the equation of state). Therefore, collapse to a black hole is not expected in this scenario; the neutron star remains stable throughout the accretion episode. In this case, two possible outcomes arise:

\begin{itemize}
\item Even in the absence of collapse, the hyperaccretion disk around an NS can power relativistic outflows \citep{zhangdai08, zhangdai09, zhangdai10}. Due to the solid surface, which prevents inward advection of heat, the inner disk becomes hotter and denser compared to the BH case. This configuration enhances neutrino cooling and can produce a jet driven by neutrino annihilation above the NS polar regions. Alternatively, if the NS is rapidly spinning and possesses an ultra-strong magnetic field ($B \gtrsim 10^{14}$ G, i.e., a magnetar), its rotational energy can directly drive a Poynting-flux dominated jet \citep{usov92, metzger11}. The maximum available energy is the magnetar's rotational energy:
\begin{equation}
E_{\mathrm{rot}} = \frac{1}{2} I \Omega^2 \approx 2\times10^{52}\,\mathrm{erg}\,
\left(\frac{P}{1\,\mathrm{ms}}\right)^{-2},
\end{equation}
where $I \sim 10^{45}\,\mathrm{g\,cm^2}$ is the moment of inertia and $P$ the spin period.

\item It is also possible that the NS companion fails to produce any significant jet, either because the accretion rate is too low or because the necessary conditions for jet formation are not met. In such systems, the GRB would appear as a standard single-jet event, indistinguishable from those originating from isolated progenitors.
\end{itemize}

Thus, the NS companion scenario encompasses a range of possibilities, from no second jet at all (ordinary LGRB) to a powerful second jet that may rival or exceed J1, with the specific outcome depending on the accretion rate, NS spin, and magnetic field strength.

\subsection{Interaction and Dynamics of the Two Jets}
\label{sec:interaction}

Observational results from LIGO/Virgo on merging binary black holes (BBHs) provide important clues about the spin orientations in close binaries. Statistical analyses of the GWTC-2 and GWTC-3 catalogs suggest that the distribution of the effective inspiral spin parameter $\chi_{\mathrm{eff}}$ is consistent with a mixture of aligned and misaligned spins, with a significant fraction of systems showing evidence for spin-orbit misalignment \citep{callister21, galaudage21,2023PhRvX..13a1048A,2025arXiv250818083T}. This implies that the spin axes of the two BHs in a merging binary are not necessarily aligned with each other or with the orbital angular momentum. Since relativistic jets are expected to be launched along the spin axes of the collapsing star (for J1) and the accreting companion (for J2), these observational findings suggest that the two jets in our scenario may also have a range of relative orientations. We denote the angle between the two jet axes as $\phi$, which can vary from nearly aligned ($\phi \approx 0$) to significantly misaligned ($\phi \gtrsim \theta_j$, where $\theta_j$ is the jet half-opening angle).

Regarding the Lorentz factor of J2, the accretion rate onto the companion is typically one to two orders of magnitude lower than that which powered J1, suggesting that the intrinsic power of J2 is generally more modest. All else being equal, this tends to yield $\Gamma_2 < \Gamma_1$. However, one important effect can modify this expectation: if J2 propagates inside the cavity cleared by J1 along the polar axis, mass loading from the surrounding ejecta is minimal, allowing $\Gamma_2$ to approach its intrinsic value. In such cases, $\Gamma_2$ could potentially be comparable to or even slightly exceed $\Gamma_1$, depending on the details of the accretion flow and jet launching efficiency. For the purpose of this work, we treat $\Gamma_2/\Gamma_1$ as a free parameter and consider both possibilities in the dynamical analysis below.

When the two jets are approximately aligned, their subsequent evolution depends on the Lorentz factor ratio. If $\Gamma_2 < \Gamma_1$, J2 never catches up with the head of J1. However, as J1 propagates outward, it eventually decelerates upon sweeping up enough circumstellar medium. When the velocity of J1 decreases to become comparable to $\Gamma_2$, J2 effectively catches up and attaches to the tail of J1, injecting its energy into the leading outflow. {This process is analogous to the energy injection model often invoked to explain GRB afterglow plateaus \citep{rees98,sari01},}  but here the energy source is a distinct second engine rather than prolonged activity of a single central engine. The combined outflow then continues to propagate outward, with the injected energy modifying the subsequent afterglow dynamics and evolution.

The deceleration radius of J1, where it has swept up enough circumstellar medium to begin significant slowing, is given by \citep[e.g.,][]{gao13}
\begin{equation}
R_{\mathrm{dec,1}} \approx \left( \frac{3E_{\mathrm{iso,1}}}{4\pi n m_p c^2 \Gamma_1^2} \right)^{1/3},
\end{equation}
where $E_{\mathrm{iso,1}} \sim 10^{52}$--$10^{53}\,\mathrm{erg}$ is the isotropic-equivalent energy of J1 and $n \sim 1\,\mathrm{cm}^{-3}$ is the typical circumstellar density. For $\Gamma_1 \sim 100$, this yields $R_{\mathrm{dec,1}} \sim 10^{16}$--$10^{17}$ cm.

If $\Gamma_2 > \Gamma_1$, J2 can catch up from behind and the collision radius is approximately $R_{\mathrm{coll}} \approx 2\Gamma_1^2 c\,\Delta t_{\mathrm{delay}}$. For typical parameters ($\Gamma_1 \sim 100$, $\Delta t_{\mathrm{delay}} \sim 10$--$100$ s), we obtain $R_{\mathrm{coll}} \sim 10^{15}$--$10^{16}$ cm. Comparing with the deceleration radius, we find $R_{\mathrm{coll}} \lesssim R_{\mathrm{dec,1}}$, suggesting that the collision occurs while J1 is still in its coasting phase. This ensures that the collision involves the fast head of J1 rather than its decelerated tail, maximizing the kinetic energy available for dissipation. The collision drives a pair of relativistic shocks: a forward shock propagating into J1 and a reverse shock propagating back into J2. The collision can be classified as either violent or mild based on whether a strong shock forms at the interface between the two jets. Two conditions are necessary for violent shock formation: (1) the second jet must move supersonically relative to the first, and (2) it must carry sufficient energy to significantly heat the leading jet. Quantitatively, these criteria can be expressed as \citep{zhang02}:
\begin{equation}
\Gamma_{12} \ge 1.22, \quad \text{and} \quad (4\Gamma_{12}+3)(\Gamma_{12}-1) > \frac{4E_1}{E_2}\left[\min\left(1,\frac{R}{R_{\mathrm{s,2}}}\right)\right]^{-1},
\end{equation}
where $\Gamma_{12} = (\Gamma_1/\Gamma_2 + \Gamma_2/\Gamma_1)/2$ is the relative Lorentz factor between the two jets, {and $R_{\mathrm{s,2}}=\Gamma_2^2\Delta_2$ is the spreading radius of J2, where $\Delta_2$ denotes the thickness of J2.} When these conditions are satisfied, the collision produces a pair of relativistic shocks, a forward shock propagating into J1 and a reverse shock into J2, analogous to the internal shock mechanism in prompt GRB emission. This process efficiently dissipates kinetic energy and can generate a strong high-energy signal extending into the gamma-ray band. When the conditions are not met, the interaction is mild, resulting in gradual energy transfer without strong shock formation. {It is also worth noting that the interaction outcome depends on the composition of the jets: for matter-dominated jets, a violent collision can occur, leading to strong internal shocks \citep{kobayashi97}; for Poynting-flux dominated jets, the interaction typically results in gradual energy injection rather than strong shock formation, as the magnetic energy is not easily dissipated by shocks \citep{lyubarsky09, komissarov09, zhang11}.}

When the two jets are not aligned, they propagate independently through the circumstellar medium. Their afterglow blast waves expand laterally and eventually meet at radii comparable to the afterglow scale. The encounter of the two blast waves generates complex hydrodynamical structures, including shocked regions and possible turbulence, and can lead to characteristic morphological features such as double-lobed or X-shaped remnants.

\section{Observational signatures}
\label{sec:signatures}

Having established the dynamical regimes of the two-jet system, we now discuss the observable consequences that distinguish each scenario. The predicted signals depend critically on the Lorentz factor ratio $\Gamma_2/\Gamma_1$ and the alignment angle $\phi$ between the two jets.

\begin{figure}
\centering
\label{fig}
   \includegraphics[width=7in]{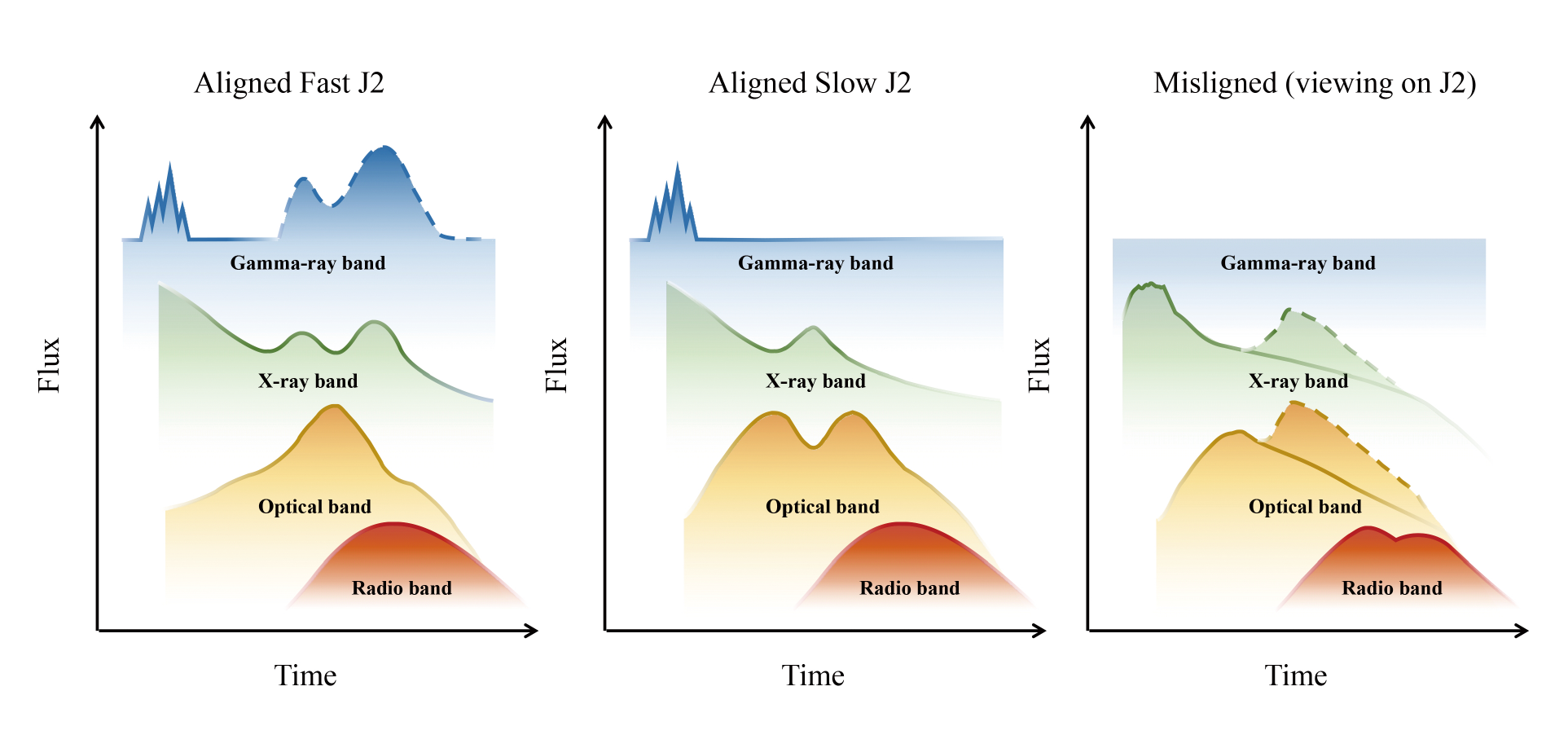}
      \caption{Schematic light curves for different two-jet scenarios. Different colors represent different observing bands, and dashed lines indicate possible components depending on specific parameters. Aligned fast jet $(\Gamma_2>\Gamma_1)$: three emission episodes (J1 prompt, J2 internal, jet-jet collision) may appear as multiple gamma-ray pulses. Aligned slow jet $(\Gamma_2<\Gamma_1)$: J2 internal emission appears as an X-ray flare, followed by an afterglow plateau from energy injection. Misaligned observer aligned with J2: an initial fast X-ray transient (FXT) is followed by late-time off-axis rebrightening from J1.}
            \end{figure}

\subsection{Aligned Fast Second Jet Scenario}
In this regime, where the second jet is aligned with and faster than the first, the observational signature may consist of three distinct radiation components. The first jet (J1) produces its prompt gamma-ray emission through internal dissipation at a typical radius $R_{\rm emit}\lesssim10^{15}$ cm, with an observer-frame duration $\delta t_1\sim10$-$100$ s and isotropic luminosity $L_{\rm iso,1}\sim10^{50}$-$10^{52}\,\rm erg\,s^{-1}$ \citep{zhang18book}. 

The second jet (J2) is launched after a physical delay $\Delta t_{\rm delay}\sim d/v_{\rm ej}$, and produces its own internal dissipation emission at a similar radius. This second pulse appears after a time interval comparable to $\Delta t_{\rm delay}$ (typically $\sim100$-$1000$ s), either as a gamma-ray re-trigger or as an X-ray flare lasting $\sim100$-$1000$ s. Shortly thereafter (with negligible additional delay), the two jets collide and their interaction depends on whether a violent shock forms at the interface. 

Violent collisions efficiently dissipates kinetic energy and can generate a strong high-energy signal extending into the gamma-ray band. The duration of such collision-induced emission in the observer frame is governed by the reverse shock crossing time of the J2 shell,
\begin{equation}
\delta T_{\rm coll}\approx \frac{\Delta_2}{c}\cdot\frac{\Gamma_2^2}{\Gamma_{\rm coll}^2(\Gamma_2/\Gamma_1-1)},
\end{equation}
where $\Gamma_{\rm coll}\sim\sqrt{\Gamma_1\Gamma_2}$ represents the Lorentz factor of the shocked region. This timescale is of the same order as J2's internal emission duration.

When the interaction is mild (e.g., for Poynting-flux dominated jets where strong shocks do not readily form), the collision results in gradual energy transfer rather than violent dissipation. In such cases, the emission appears as a broader, softer bump rather than a sharp gamma-ray pulse, effectively manifesting as energy injection into the outflow.

In this scenario, multiple gamma-ray triggers separated by $\sim100$-$1000$ s of quiescence are possible. The distinct components arise from different dissipation processes with potentially different magnetic field configurations, leading to significant evolution in spectral hardness and polarization across pulses. In the afterglow phase, the collision injects energy before J1 decelerates, refreshing the forward shock and producing a faster rise and brighter peak than in single-jet cases. If the collision or J2 internal emission occurs at lower energies (X-ray/optical), they appear as large bumps superimposed on the rising afterglow.

\subsection{Aligned Slow Second Jet Scenario}

When the second jet is aligned with but slower than the first, the first jet (J1) produces its prompt gamma-ray emission as discussed above. The second jet (J2) is launched after a physical delay $\Delta t_{\rm delay}\sim d/v_{\rm ej}$ and produces its own internal dissipation emission, but due to its lower Lorentz factor and likely lower power, this emission is expected to peak at softer energies, probably appearing as an X-ray flare lasting $\sim100$-$1000$ s rather than a gamma-ray trigger.

As described in Section~\ref{sec:interaction}, J2 never catches the head of J1 but eventually merges with its decelerated tail, injecting energy into the leading outflow. In this case, no gamma-ray pulse is produced at the moment of merger. Instead, the energy injection manifests in the afterglow phase: the forward shock is refreshed, producing a prolonged plateau or a slow-rise bump in the X-ray and optical light curves, typically on timescales of $10^3$-$10^4$ seconds. The injected energy also modifies the subsequent decay slope, potentially producing a shallower decline than expected from the standard forward shock model. 

Thus, the overall light curve in the slow-jet scenario may consist of: (i) J1 prompt gamma-ray emission, (ii) an X-ray flare from J2 internal dissipation after $\sim100$-$1000$ s, and (iii) a broad afterglow bump or plateau from energy injection, peaking at $10^3$-$10^4$ s. Unlike the fast second jet case, no sharp spectral or polarization jumps are expected at the merger time, as the interaction is gradual rather than violent.

\subsection{Misaligned Scenario}

When the two jets are misaligned, the observed signal depends critically on the observer's line of sight relative to each jet axis. Two distinct observational cases arise.

If the observer is aligned with J1, a normal GRB is detected, with its prompt emission and early afterglow dominated by J1. The prompt emission from J2, being off-axis, is strongly deboosted and unlikely to be detectable. However, the late-time radio afterglow may reveal the presence of the second jet. As both jets expand laterally into the circumstellar medium, their afterglow emission becomes increasingly isotropic. The radio emission may therefore exhibit two distinct components, potentially leading to a late-time rebrightening when J2's emission enters the line of sight. Moreover, as the two blast waves expand and eventually interact, they can produce characteristic X-shaped radio lobes similar to those observed in some giant radio galaxies\footnote{{Point-symmetric morphologies, including X-shaped structures, can also arise from the jittering jets explosion mechanism (JJEM) in core-collapse supernovae, where many stochastic jets shape the supernova ejecta on stellar scales ($\sim10^{13}$,cm) \citep{soker24, braudo25}. Our model predicts large-scale ($\gtrsim10^{17}$,cm) radio lobes from the GRB afterglow, which are produced by the interaction of two stable, relativistic jets from distinct compact objects and are observable years after the burst. The two scenarios thus differ in physical origin, scale, and observational context.}} \citep{2022A&A...660L..10D}. Such morphological features, detectable with high-resolution radio interferometry (e.g. SKA), would provide unambiguous evidence for a misaligned second jet.

If the observer is aligned with J2 instead of J1, a very different observational picture emerges. The prompt emission from J2 is detected, but due to its likely lower Lorentz factor and power, it may appear as an X-ray flash or a fast X-ray transient (FXT) rather than a classical GRB. Such events are characterized by soft spectra and durations of hundreds to thousands of seconds \cite[][and reference therein]{2022A&A...663A.168Q}. Following this X-ray trigger, the afterglow of J2 develops normally. However, if the misalignment angle is not too large, the much more energetic J1 may eventually become visible as its beaming cone widens during deceleration. This produces a late-time rebrightening in the afterglow, potentially dominating over J2's fading emission. The resulting light curve would thus consist of: (i) an initial X-ray flash from J2, (ii) a subsequent afterglow phase from J2, and (iii) a late, bright rebrightening when J1's off-axis afterglow enters the line of sight. Such behavior has been observed in some FXTs with complex afterglow rebrightenings, {e.g., EP240414a \citep{2025NatAs...9.1073S}; EP241021a \citep{2025A&A...701A.225B,2025ApJ...991..115W}.} As in the previous case, late-time radio imaging may reveal X-shaped lobe structures from the two interacting blast waves, providing a definitive test of the misaligned geometry.

\section{Conclusions and Discussion}
\label{sec:conclusion}

Through the above analysis, we have developed a comprehensive model for long gamma-ray bursts originating from close binary systems where the companion is a compact object (black hole or neutron star). We find that such progenitor systems can produce a variety of distinctive observational signatures, including: (i) intermittent gamma-ray triggers separated by quiescent periods of $\sim100$-$1000$ s; (ii) normal LGRBs followed by delayed X-ray flares and optical bumps; (iii) fast X-ray transients as the initial trigger (when the line of sight aligns with J2), with subsequent late-time rebrightening from off-axis J1 afterglow; (iv) late-time radio structure such as X-shaped lobes or rebrightening features from the interaction of two misaligned blast waves; and (v) complex afterglow plateaus and bumps from energy injection in the slow second jet.

These predicted phenomena have indeed been observed in some GRBs. For instance, long quiescent periods in prompt emission have been seen in events like GRB 220627A, which exhibits a double-pulse profile with a $\sim15$ minute quiescent interval and different spectral energy distributions between the two pulses, disfavoring a gravitational lensing origin and suggesting a multi-component jet \citep{2022ApJ...940L..36H,2023A&A...677A..32D}. An even more extreme case is GRB 250702B, which shows multiple distinct triggers separated by exceptionally long quiescent intervals \citep{zhang26}. Furthermore, systematic studies have revealed a class of GRBs exhibiting giant X-ray and optical bumps with rapid rises followed by long decays (e.g., GRBs 121027A and 111209A; \citealt{zhao20}). Whether these specific events arise from the double-jet mechanism will require detailed, case-by-case modeling in future work. 

The observational features predicted in our model share some similarities with the fallback accretion model, which has been invoked to explain X-ray flares and late-time energy injection \citep{wu13,gao16,zhao20}. In the fallback scenario, material from the progenitor star that failed to escape during the initial explosion eventually falls back toward the central black hole, reactivating the engine and producing a characteristic light curve often with a smooth, gradual rise and decay following a broken power-law form, and a typical $t^{-5/3}$ decay at late times \citep{wu13,gao16}. In contrast, our double-jet model can produce sharper, multi-peaked structures from distinct dissipation events, and most distinctively, it predicts the possibility of spatially resolved X-shaped or double-lobed radio structures from the interaction of two misaligned blast waves, which cannot be produced in any single-engine model including fallback accretion. Such morphological features, detectable with high-resolution radio interferometry, would provide unambiguous evidence for the double-jet scenario.

The event rate of double-jet systems remains uncertain due to several poorly constrained factors: the efficiency of common envelope evolution, the survival probability after the first supernova, the fraction of systems where the companion successfully launches a jet, and the beaming and alignment probabilities. Recent binary population synthesis studies suggest that $\sim20\%$--$85\%$ of LGRBs arise from close binary systems that have survived the first supernova \citep{bavera22}. The distinctive signals predicted in our model, such as multiple gamma-ray triggers, delayed X-ray/optical bumps, and late-time radio structures, provide a promising observational avenue to test this conclusion and constrain the uncertain parameters.

In the context of binary progenitors, the “induced gravitational collapse” (IGC) paradigm has been developed to explain long GRBs from tight CO-NS binaries \citep{rueda12, fryer14, becerra15, becerra16}. In this model, the supernova ejecta from the exploding CO core trigger hypercritical accretion onto the NS companion, causing it to collapse to a black hole; the newly formed BH then powers a single GRB jet. Our double-jet scenario differs in two key aspects: (i) the compact companion is already a black hole (or a neutron star that does not collapse, given the estimated fallback rates), and (ii) the first jet originates from the primary core collapse, while the second jet is launched by the companion’s accretion disk. This naturally leads to two distinct relativistic outflows whose interactions produce the rich multi-component phenomenology discussed in this work, rather than a single jet from a newly formed BH.

{Beyond the IGC framework, other mechanisms can also produce complex light curves, such as jet axis variability from a single engine via the wobbling mechanism \citep{gottlieb22}. However, our double-jet scenario differs in several key aspects: (i) the time delay between episodes is set by the binary separation ($\sim100$–$1000$ s), much longer than the wobbling timescale; (ii) the second jet may peak at X-ray energies, while in the wobbling case all emission is expected in gamma-rays; (iii) the second jet’s duration is also longer; and (iv) the afterglow signatures provide a potential discriminant. In particular, non-axisymmetric structured jets (including those from wobbling) are predicted to produce complex afterglow light curves with multiple peaks or plateaus \citep{li23}, as well as distinctive polarization evolution characterized by significant fluctuations in polarization degree and rotation of the polarization angle \citep{lijd25}. Future multi-wavelength and polarimetric observations may thus help distinguish between the two scenarios.}

Distinguishing these scenarios will require high-quality data from upcoming facilities. High-resolution radio interferometry with \textit{SKA} \citep{2009IEEEP..97.1482D} can resolve X-shaped lobe structures in late-time afterglows. Time-resolved polarimetry with \textit{POLAR-2} \citep{2021SPIE11444E..2VH} and \textit{eXTP} \citep{zhang19} can detect sharp polarization angle jumps characteristic of jet-jet collisions. Multi-wavelength time-domain surveys with \textit{Einstein Probe} \citep{yuan25} and \textit{SVOM} \citep{wei16} will provide larger samples of GRBs with complex light curves for statistical studies. Ultimately, confirming the double-jet scenario would provide valuable constraints on the evolution of close massive binaries and their role as progenitors of both GRBs and gravitational wave sources.

\begin{acknowledgments}
We thank Bing Zhang, Weihua Lei, Xingjiang Zhu and Yin Qin for their helpful discussions and comments. We thank Xudong Wen for his assistance with the cartoon illustration. We thank the anonymous referee for the helpful comments. This work was supported by the National Natural Science Foundation of China (Projects 12541304, 12373040), the National Key R\&D Program of China (grant Nos. 2024YFA1611703 and 2024YFA1611700)
and the Fundamental Research Funds for the Central
Universities.
\end{acknowledgments}

\bibliography{sample701}{}
\bibliographystyle{aasjournalv7}



\end{document}